\documentclass[jkps,preprint]{revtex4}
\usepackage{graphicx}
\usepackage{amssymb}
\usepackage{amsmath}
\usepackage{bm}

\usepackage{xcolor}

\begin{document}
\setcounter{page}{0}
\title[]{Ermakov-Lewis invariant in single field inflation}
\author{Seoktae \surname{Koh}}
\email{kundol.koh@gmail.com}
\affiliation{ Department of Science Education, Jeju National University, Jeju, 63243, Korea}

\date[]{}

\begin{abstract}
We investigate the dynamical symmetry of the Sasaki-Mukhanov equation, which governs the evolution of primordial perturbations during inflation. By mapping the Sasaki-Mukhanov equation to a constant-frequency harmonic oscillator via an auxiliary Ermakov field satisfying the Ermakov-Pinney equation, we construct the associated $sl(2, R)$ generators and derive the Ermakov-Lewis invariant. We apply this framework to de Sitter space and slow-roll inflation, demonstrating that the Bunch-Davies vacuum gives $I_{EL} = 1/2$, conserved non-perturbatively at all orders in slow-roll.
\end{abstract}


\maketitle

\section{Introduction}
The theory of cosmological perturbations is the cornerstone of modern inflationary cosmology \cite{Bardeen:1980kt, Kodama:1984ziu}. At the linear level, the evolution of scalar perturbations is governed by the Sasaki-Mukhanov (SM) equation \cite{Sasaki:1986hm,Mukhanov:1988jd}:
\begin{align}
    v_k'' + \left( k^2 - \frac{z''}{z}\right) v_k = 0,
\end{align}
where the prime denotes the derivative with respect to the conformal time $\eta$ and $k$ is a wave vector. This equation implies that quantum fluctuations during the inflationary epoch are stretched to super-horizon scales \cite{Starobinsky:1979ty,Guth:1980zm,Linde:1981mu}, ultimately seeding the Large Scale Structure (LSS) and the Cosmic Microwave Background (CMB) anisotropies \cite{Planck:2018jri}.

A critical observation is that the Sasaki-Mukhanov equation takes the exact form of a time-dependent harmonic oscillator  \cite{Lewis:1967, Lewis:1968tm}. The term $\omega^2(\eta) = k^2 - \frac{z''}{z}$ acts as a time-varying frequency squared. Unlike a standard oscillator with constant frequency, the effective mass term $\frac{z''}{z}$ depends on the background dynamics of the universe \cite{Mukhanov:1990me}. This time-dependence is responsible for particle production \cite{Parker:1969au}  and the squeezing of the initial vacuum state \cite{Grishchuk:1990bj, Albrecht:1992kf, Polarski:1995jg}.

The symmetry of a standard harmonic oscillator is described by the $SL(2, \mathbb{R})$ group (isomorphic to $SU(1,1)$) \cite{Wybourne:1974}. However, for a time-dependent harmonic oscillator, this symmetry is broken by the explicit time-dependence of $\omega(\eta)$. To restore a dynamical symmetry, we introduce an auxiliary Ermakov field $\rho(\eta)$ satisfying the Ermakov-Pinney equation \cite{Ermakov:1880, Pinney:1950}
\begin{align}
	\frac{d^2 \rho}{d\eta^2} + \omega^2(\eta)\rho = \frac{\Omega^2}{\rho^3}
\end{align}
Using this auxiliary function, we construct the Ermakov-Lewis invariant $I_{EL}$\cite{Lewis:1967, Lewis:1968tm, Leach:1977a, Leach:1977fa}, which remains constant under the full evolution of the Sasaki-Mukhanov equation. The three generators of the $sl(2,R)$ algebra constructed from the Ermakov-transformed variables provide a complete algebraic characterization of the system.

In this work, we apply this algebraic framework to inflationary scenarios. We first consider pure de Sitter Space,  where $z''/z = 2/\eta^2$, allowing for exact solutions of the Ermakov-Pinney equation. We then extend the analysis to slow-roll inflation \cite{Liddle:1992wi, Lidsey:1995np}, where the slow-roll parameters $\epsilon$ and $\delta$ shift the index $\nu$ away from $3/2$, examining how the squeezing generators $T_1$ and $T_2$ capture the departures from exact scale invariance while $I_{EL} = 1/2$ remains conserved non-perturbatively. 

\section{Ermakov systems of the Sasaki-Mukhanov equation} \label{elsystem}

The quadratic action for scalar perturbations during inflation is given by
\begin{align}
   S = \int d^3k d\eta \frac{1}{2}\Big[{v_k'}^2 - \omega_k^2(\eta) v_k^2 \Big]
\end{align}
where $v_k(\eta) = z\mathcal{R}_k$ is the canonical Sasaki-Mukhanov variable  \cite{Sasaki:1986hm, Mukhanov:1988jd}, $\mathcal{R}_k$ is the gauge-invariant curvature perturbation, and $z$ is the pumping field. The time-dependent frequency $\omega_k(\eta)$ for Einstein gravity minimally coupled to a canonical scalar field is
\begin{align}
    \omega_k^2 (\eta) = k^2 - \frac{z''}{z},
\end{align}
with $z = a \dot{\phi}/H$. The equation of motion derived form this quadratic action is the Sasaki-Mukhanov equation
\begin{align}
    v_k'' + \left( k^2 - \frac{z''}{z}\right) v_k = 0,	
    \label{smeq}
\end{align}
which has the form of a time-dependent harmonic oscillator for each Fourier mode $k$, with a time-dependent frequency $\omega_k(\eta)$. 

To analyze the dynamical structure of Eq. (\ref{smeq}), we introduce an auxiliary real positive-definite function $\rho_k(\eta)$ and define a new variable $u_k$ and a new time coordinate $\tau$ by
\begin{align}
    u_k = v_k/\rho_k (\eta), \quad \tau \equiv \theta_k(\eta) = \int^{\eta} \frac{1}{\rho_k^2 (\eta')} d\eta',
    \label{transf}
\end{align}
Under this transformation, Eq. (\ref{smeq}) becomes
\begin{align}
    \frac{d^2 u_k}{d\tau^2} + \Omega^2 u_k = 0,
    \label{tihoeq}
\end{align}
a simple harmonic oscillator with a constant frequency $\Omega$, provided that  $\rho_k(\eta)$ satisfies the nonlinear Ermakov-Pinney equation \cite{Ermakov:1880, Pinney:1950}
\begin{align}
    \rho_k'' + \omega^2 (\eta) \rho_k = \frac{\Omega^2}{\rho_k^3},
    \label{epeq}
\end{align}
where $\Omega^2$ is an arbitrary positive constant, encoding the residual freedom in the choice of $\rho_k$. Sasaki-Mukhanov equation (\ref{smeq}) and Ermakov-Pinney equation (\ref{epeq}) together consist of the Ermakov systems \cite{Leach:1977a, Leach:1977fa}. Since the solution to (\ref{tihoeq}) is simply $u_k = e^{\pm i\Omega \tau}$, the corresponding solution to the original Sasaki-Mukhanov equation (\ref{smeq}) is 
\begin{align}
    v_k = \frac{\rho_k(\eta) }{\sqrt{2}} e^{\pm i\Omega \theta_k(\eta)}. 	
    \label{modslogen}
\end{align}
which can be recognized as a generalized WKB-type solution governed entirely by the Ermakov amplitude $\rho_k$ and phase $\theta_k$.
The Wronskian of two independent solution is 
\begin{align}
    W = v_k {v_k^*}' - v_k' v_k^* =  \mp i \Omega\rho_k^2 \theta_k' = \mp i\Omega = \text{const.},
    \label{wronskian}
\end{align}
where we have used $\rho_k^2 \theta_k' = 1$ from (\ref{transf}). For the conventional Bunch-Davies normalization one sets $\Omega = 1$, so that $W = \mp i$.

In the transformed coordinate $\tau$, the system (\ref{tihoeq}) is an autonomous harmonic oscillator with constant frequency $\Omega$, whose Hamiltonian is therefore a constant of motion
\begin{align}
    H_k (u_k,\pi_k; \tau) = \frac{1}{2} \pi_k^2 + \frac{1}{2} \Omega^2 u_k^2 \equiv \text{const}.,	\label{hamilu}
\end{align}
where the canonical momentum conjugate to $u_k$ with respect to $\tau$ is
\begin{align}
    \pi_k = \frac{\partial u_k}{\partial \tau} = \rho_k v_k' - \rho_k' v_k. 	
\end{align}
Expressing the Hamiltonian (\ref{hamilu}) back in terms of the original variable $v_k$ and conformal time $\eta$, one obtains 
\begin{align}
    I_k \equiv H_k (u,\pi;\tau) = \frac{1}{2} \left[ (\rho_k v_k' - \rho_k' v_k)^2 +\Omega^2 \left(\frac{v_k}{\rho_k}\right)^2 \right],	
\end{align}
which is the Ermakov-Lewis invariant \cite{Lewis:1967, Lewis:1968tm} of the original system. The conservation of this quantity, $\frac{dI_k}{d\eta} = 0$, follows directly from the fact that $I_k$ equals the energy of the constant-frequency oscillator (\ref{tihoeq}), which is time-independent in $\tau$.

The general solution to the Ermakov-Pinney equation (\ref{epeq}) was given by Pinney \cite{Pinney:1950} as
\begin{align}
    \rho_k(\eta) = 	\sqrt{A v_1^2 + 2B  v_1 v_2 + C v_2^2} \label{rhosol}
\end{align}
where $v_1$ and $v_2$ are two linearly independent solutions of (\ref{smeq}), $W_v \equiv W(v_1, v_2)$ is their Wronskian, and the real constants $A,\,B,\,C$ satisfy $AC-B^2 = \Omega^2/W_v^2$. This solution makes explicit how the Ermakov amplitude $\rho_k$ inherits its full time dependence from the background geometry through $v_1$ and $v_2$, while the invariant $I_k$ remain exactly conserved throughout the evolution.

For a single time-dependent harmonic oscillator, this construction is completely general. For any $\omega(\eta)$, two independent solutions of the time-dependent harmonic oscillator furnish an Ermakov amplitude $\rho$ solving the Pinney equation, and the corresponding Ermakov-Lewis invariant exists for arbitrary $\omega(\eta)$, with no restriction on its functional form.

This construction holds even at the quantum level. The quantum operator $\hat{I}_{E L}$ satisfies the Liouville-von Neumann equation exactly \cite{Lewis:1968tm}, 
\[
\frac{\partial \hat{I}_k}{\partial \eta} + \frac{1}{i\hbar} [\hat{I}_k,\, H] = 0,
\]
by which the eigenvalues of $\hat{I}_k$ are time-independent and its  eigenstates are exact solutions of the time-dependent Schr\"odinger equation. No quantum correction spoils the invariant equation, and the Ermakov-Lewis construction remains exact after quantization.

\section{SL(2,R) symmetry of harmonic oscillator}

We investigate the Ermakov-Lewis invariant derived in Section \ref{elsystem} as a manifestation of the $SL(2,\mathbb{R})$ dynamical symmetry underlying the time-dependent harmonic oscillator. 

The harmonic oscillator with unit frequency $V(Q) = \frac{1}{2}Q^2$, possesses a hidden $SL(2,\mathbb{R})$ symmetry analogous to that of the free particle and the conformal oscillator \cite{deAlfaro:1976vlx,Niederer:1973tz}, realized through the time-dependent canonical transformations \cite{Niederer:1973tz}
\begin{align}
    \delta t  = \epsilon f(t), \quad \delta Q = \frac{1}{2} \epsilon \dot{f}(t) Q,	
\end{align}
where $\epsilon$ is an infinitesimal parameter and $f(t)$ is constrained by the requirement that the action be invariant. Substituting into the equations of motion one finds
\begin{align}
    \dddot{f} + 4 \dot{f} = 0,
\end{align}
whose general solution is the three-parameter family
\begin{align}
    f(t) = c_1 + c_2 \sin 2t + c_3 \cos 2t.	
\end{align}
The three linearly independent modes correspond, via N\"oether's theorem, to three conserved charges. For $f = 1$ one recovers the Hamiltonian; the oscillatory modes generate two additional invariants:
\begin{align}
    \mathcal{C}_1 =& 2 T_0 = H,
    \\
    \mathcal{C}_2 =& T_1\, \sin 2t - T_2 \,\cos 2t, 
    \label{genho1}
    \\
    \mathcal{C}_3 =& T_1\, \cos 2t + T_2\, \sin 2t , \label{genho2}
\end{align}
where the three fundamental, time-independent phase-space generators are
\begin{align}
    T_0 = \frac{1}{4} (P^2 + Q^2), \quad T_1 = \frac{1}{4} (P^2 - Q^2), \quad T_2 = \frac{1}{2} PQ.	\label{genho}
\end{align}
These generators satisfy the standard $sl(2,R)$ Lie algebra
\begin{align}
    \{T_0,\, T_1\} = T_2, \,\, \{T_1,\, T_2\} = - T_0, \,\, \{T_2,\, T_0\} = T_1.	\label{sl2rso}
\end{align}
The time-dependent charges $\mathcal{C}_2$ and $\mathcal{C}_3$ are simply rotations of the generators $T_1$ and $T_2$ in the $(T_1,T_2)$ plane of $sl(2, R)$.

The time-dependent harmonic oscillator
\begin{align}
    \ddot{Q} + \omega^2(t) Q = 0,	\label{tdhoeq}
\end{align}
retains an $SL(2, \mathbb{R})$ dynamical symmetry for any $\omega(t)$. However, for a general time-dependent frequency, the generators can no longer be expressed in the closed trigonometric forms of Eqs. (\ref{genho1}) and (\ref{genho2}), they must instead be constructed from the solutions of the Ermakov-Pinney equation (\ref{epeq}). 
Using the rescaled variable and time introduced in (\ref{transf}),
\begin{align}
    Q_c = \frac{Q}{\rho(t)}, \quad \tau = \theta(t) = \int^{t'} \frac{1}{\rho^2(t')} dt',	
\end{align}
the Lagrangian transforms to that of a constant-frequency oscillator,
\begin{align}
    Ldt = \tilde{L}d\tau  = \frac{1}{2} \Big( Q_c'^2 - \Omega^2 Q_c'^2 \Big)d\tau	
\end{align}
provided $\rho(t)$ satisfies the Ermakov-Pinney equation \cite{Ermakov:1880, Pinney:1950} 
\begin{align}
    \ddot{\rho} +\omega^2 (t) \rho = \frac{\Omega^2}{\rho^3}.	\label{epeq2}
\end{align}
The conjugate momentum in the new coordinate is
$P_c = \rho P - \dot{\rho} Q$.

The general solution to (\ref{epeq2}) is given by Eq. (\ref{rhosol})
where $v_1$ and $v_2$ are any two linearly independent solutions of the time-dependent harmonic oscillator (\ref{tdhoeq}) with Wronskian $W_v$, and the constants $A,\,B,\,C$ satisfy $AC-B^2 = \Omega^2/W_v^2$. Writing (\ref{rhosol}) $ \rho^2 = {\bf v}^T \mathbf{M} {\bf v}$ with 
\begin{align}
    \mathbf{v} =  \begin{pmatrix} v_1\\ v_2 \end{pmatrix}, \quad \mathbf{M} = \begin{pmatrix} A & B \\ B & C \end{pmatrix}, \quad \det \mathbf{M} = \frac{\Omega^2}{W_v^2},
\end{align}
the three independent entries of $\mathbf{M}$ correspond precisely to the three parameters of $SL(2,\mathbb{R})$. This implies that different choices of $(A,B,C)$ with fixed $\det \mathbf{M}$ correspond to different choices of basis for the $sl(2,R)$ invariants,  {\it i.e.} to different representations of the same dynamical algebra. 

Under a change of basis of the solution space, $\mathbf{v} \rightarrow \mathbf{v}' = \Lambda \mathbf{v}$ with $\Lambda \in SL(2, \mathbb{R})$, the invariance of $\rho^2$ requires $\mathbf{M}$  to transform as
\begin{align}
    \mathbf{M}' = (\Lambda^{-1})^{T} \mathbf{M} \Lambda^{-1}	
\end{align}
from which  $\det \mathbf{M}' = \det \mathbf{M}$ follows immediately  since $\det \Lambda = 1$.
This confirms that the Wronskian is preserved under the full $SL(2, \mathbb{R})$  group: $W_v' = W_v$

In the transformed coordinate system ($\tau, Q_c, P_c$), the dynamics is that of a simple harmonic oscillator with $\Omega =1$. With the standard  Bunch-Davies canonical normalization $W = \pm i$ and constructing the $sl(2,R)$ generators by direct analogy with Eq. (\ref{sl2rso}), we obtain
\begin{align}
    T_0 = \frac{1}{4} (P_c^2 + Q_c^2) = \frac{1}{2} I_{EL}, \quad T_1 = \frac{1}{4} (P_c^2 - Q_c^2), \quad T_2 = \frac{1}{2} Q_c P_c,
\end{align}
which satisfy the same algebra (\ref{sl2rso}). Here $I_{EL} =2 T_0$ is identified as the Ermakov-Lewis invariant, which in the original variable reads
\begin{align}
    I_{EL} = 	\frac{1}{2} \left[(\rho P - \dot{\rho} Q)^2 + \left(\frac{Q}{\rho} \right)^2 \right].
\end{align}

A rotation by the phase $\theta(t)$ in the $(T_1, T_2)$ sub-plane of $sl(2,R)$ yields the complete set of three conserved charges
\begin{align}
    I_{EL} =&\,\,  2 T_0,
    \\
    I_1 =&\,\,  T_1 \cos 2\theta + T_2 \sin 2\theta, \label{gentdho1}
    \\
    I_2 =&\,\, - T_1 \sin 2\theta + T_2 \cos 2\theta. \label{gentdho2}
\end{align}
All three are exact, non-perturbative constants of motion for any realization of $\omega(t)$. Physically, $I_{EL}$ measures the total action in the Ermakov phase space; $I_1$ and $I_2$ encode the squeezing amplitude and squeezing angle of the quantum state, respectively, and are directly related to the Bogoliubov coefficients connecting the initial and final mode functions.

\section{Application to single scalar field inflation}

We now apply the algebraic framework developed in previous sections to concrete inflationary backgrounds. We treat first the exactly solvable case of pure de Sitter space, then slow-roll inflation as a perturbative deformation.

\subsection{de Sitter}
In exact de Sitter spacetime, the scale factor in conformal time is  $a = - 1/H\eta$\,\, with  $-\infty < \eta < 0$ and $H=\text{const.}$ With $z = a\dot{\phi}/H $, we have
\begin{align}
    \frac{z''}{z} = \frac{\nu^2 - 1/4}{\eta^2},\quad \nu = \frac{3}{2}. 
\end{align}
so the equation (\ref{smeq}) reduces to a Bessel equation of order $\nu=3/2$. Two real, linearly independent solutions are
\begin{align}
    v_1(\eta) = \sqrt{-\eta} J_{3/2} (-k\eta), \,\, v_2 (\eta) = \sqrt{-\eta} Y_{3/2} (-k\eta) 	
\end{align}
with Wronskian $W(v_1,v_2) = v_1 v_2' - v_1' v_2 = 2/\pi$, which follows directly from the Bessel identity $W(J_{\nu}, Y_{\nu})(z) = 2/(\pi z)$. 

The Bunch-Davies mode function \cite{Bunch:1978yq}, which reduces to a positive-frequency plane wave in the sub-horizon limit $k|\eta| \gg 1$,  is the analytic combination
\begin{align}
	v_k^{BD}(\eta) = \frac{\sqrt{\pi}}{2}\sqrt{-\eta}\, H_{3/2}^{(1)}(-k\eta) 
    \label{smsolwbd}
\end{align}
where $H_{\nu}^{(1)} = J_{\nu} + i Y_{\nu}$ is the Hankel function of the first kind. Using the identity
\begin{align}
    H_{\nu}^{(1)}(z) H_{\nu}^{(2)'}(z) - H_{\nu}^{(1)'} (z) H_{\nu}^{(2)} (z) = \frac{-4i}{\pi z}, 
\end{align}
the Wronskian of the Bunch-Davies mode function and its complex conjugates evaluates to
\begin{align}
   W = v_k^{BD} {v_k^{BD *}}' - {v_k^{BD}}' v_k^{BD *} = i.
\end{align}

With the symmetric choice of the coefficient $A = C = \pi/2,\,\, B=0$, which satisfies the constraint $AC-B^2 = \Omega^2/W_v^2 =\pi^2/4$ for $\Omega=1$ and $W_v = 2/\pi$, the Pinney solution (\ref{rhosol}) yields
\begin{align}
	\rho (\eta) = \sqrt{\frac{\pi}{2k}} \sqrt{-k\eta}\, \left|H^{(1)}_{3/2}(-k\eta)\right| = \sqrt{2} \left|v_k^{BD} \right|.
\end{align}

The Ermakov-Lewis invariant evaluated on the Bunch-Davies solution is
\begin{align}
    I_{EL} = \frac{1}{2} \left[\left|\rho_k {v_k^{BD}}' - \rho_k'  v_k^{BD}\right|^2 + \left|\frac{v_k^{BD}}{\rho_k}\right|^2 \right].
    \label{elbd}
\end{align}
In the sub-horizon limit ($k\eta \gg 1$), inserting $\rho_k = k^{-1/2}$ and $v_k^{BD} = e^{-ik\eta}/\sqrt{2k}$, we obtain 
\begin{align}
    I_{EL} = \frac{1}{2} \left[\frac{1}{k} \cdot \frac{k^2}{2k} + k \cdot \frac{1}{2k} \right] = \frac{1}{2},
\end{align}
where we have used $|H^{(1)}_{\nu}(z)| \approx \sqrt{\frac{2}{\pi z}}$ for $z\gg 1$. 

For $\nu=3/2$, the Hankel function has the closed-form expression
\begin{align}
    H^{(1)}_{3/2} (z) = -\sqrt{\frac{2}{\pi z}} \left( 1 + \frac{i}{z} \right) e^{iz}, 	
\end{align}
then we obtain
\begin{align}
    \rho_k = \frac{1}{\sqrt{k}} \sqrt{1+ \frac{1}{k^2 \eta^2}}, \quad v_k^{BD} = \frac{-i}{\sqrt{2k}} \left( 1+ \frac{i}{k\eta} \right) e^{-ik\eta}.	
\end{align}
From the polar decomposition satisfying the Wronskian condition $\rho_k^2 \theta_k' =1$, which encodes the canonical normalization $W = i$, the mode function takes the form $ v_k^{BD} = \frac{\rho_k}{\sqrt{2}} e^{-i\theta_k}$ from (\ref{modslogen}),  where the sign convention for the phase is chosen to match the positive-frequency mode in the sub-horizon limit. The phase $\theta_k(\eta)$ is then given by
\begin{align}
    \theta_k (\eta) =\int \frac{d\eta}{\rho_k^2} = \int \frac{k^3\eta^2}{1+k^2\eta^2}d\eta = k\eta - \tan^{-1} (k\eta),
\end{align}
up to an irrelevant constant.
Hence, one obtains
\begin{align}
    v_k^{BD} = \frac{\rho_k}{\sqrt{2}} \exp\left[-ik\eta + i\tan^{-1}(k\eta)\right].	
\end{align}

Using the amplitude $\rho_k$ and the phase $\theta_k$, we get 
\begin{align}
    \left|\rho_k {v_k^{BD}}' - \rho_k' v_k^{BD} \right|^2 =& \frac{1}{2}, \quad \frac{|v_k^{BD}|^2}{\rho_k^2} = \frac{1}{2},	
\end{align}
from which one can immediately follows that $I_{EL} = 1/2$. Since the Ermakov-Lewis invariant is conserved, this value remains fixed throughout the entire evolution.

In particular, since the Bunch-Davies initial condition corresponds to a minimum-uncertainty state, the value $I_{EL} = 1/2$ is a direct manifestation of the quantum of action inherited from the canonical normalization $W = i$.

\subsection{Slow-roll inflation}

In slow-roll inflation the scale factor takes the quasi-de Sitter form $a(\eta) \approx -1/(H\eta)$ with $H$ slowly varying. With $z = a\dot{\phi}/H$ for the single canonical scalar field, we obtain \cite{Liddle:1992wi, Lidsey:1995np, Stewart:1993bc}  in (\ref{smeq})
\begin{align}
    \frac{z''}{z} = a^2 H^2 \left(2 -\epsilon + \frac{3}{2} \delta + \dots \right) \approx \frac{\nu^2 -1/4}{\eta^2}, \quad \nu = \frac{3}{2} + 2\epsilon -\delta + \vartheta(\epsilon^2, \delta^2, \epsilon \delta),
\end{align}
where $\epsilon = -\dot{H}/H^2,\,\,\delta = \ddot{\phi}/H \dot{\phi}$.

The Sasaki-Mukhanov equation (\ref{smeq}) again reduces to a Bessel equation with the slow-roll-corrected index $\nu$, whose solution with the Bunch-Davies vacuum  is
\begin{align}
    v_k^{BD} (\eta) = \frac{\sqrt{\pi}}{2} \sqrt{-\eta} H_{\nu}^{(1)} (-k\eta),
    \label{modsolsr}
\end{align}
identical to the de Sitter results (\ref{smsolwbd}) with the replacement $3/2 \rightarrow \nu$. 

The corresponding Ermakov amplitude is
\begin{align}
	\rho_k(\eta) = \sqrt{\frac{\pi}{2k}} \sqrt{-k\eta} 
    \left|H_{\nu}^{(1)} (-k\eta)\right|,
\end{align}
which satisfies the Wronskian condition $\rho_k^2 \theta_k' = 1$. In the sub-horizon limit the $\nu$-dependence disappears and the Ermakov-Lewis invariant remains $I_{EL} = 1/2$, independent of $\nu$ and hence independent of the slow-roll parameters. In the super-horizon limit, the mode function decomposes into the two independent asymptotic branches,
\begin{align}
	v_k^{BD} \sim &\,\, \frac{C_{\nu}}{\sqrt{k}} (-k\eta)^{1/2-\nu} + \frac{D_{\nu}}{\sqrt{k}}(-k\eta)^{1/2+\nu}, \label{smsolsrsuper}
\end{align}
where
\begin{align}
    C_{\nu} =&\,\,  \frac{-i 2^{\nu-1}\Gamma(\nu)}{\sqrt{\pi}},\,\, D_{\nu} = \frac{\sqrt{\pi}}{2^{\nu+1} \Gamma(\nu+1)}. 
\end{align}
The first term corresponds to the growing mode, while the second term is the decaying mode. Although the decaying mode is subdominant in amplitude, it is essential for preserving the Wronskian normalization and the associated Ermakov–Lewis invariant.

Accordingly, the Ermakov amplitude becomes
\begin{align}
   \rho_k \sim \frac{\sqrt{2} |C_{\nu}|}{\sqrt{k}} (-k\eta)^{1/2 -\nu} \left(1+ \frac{|D_{\nu}|^2}{2|C_{\nu}|^2} (-k\eta)^{4\nu} \right), \label{rhosolsrsuper} 
\end{align}
where only the correction associated with the independent decaying branch has been retained. For $\nu > 3/2$, additional subleading corrections arise from higher-order terms within the growing branch itself, but these do not affect the phase structure responsible for the conserved invariant.
As in the de Sitter case, the phase is obtained from the polar decomposition of $v_k^{BD}$,
\begin{align}
    \theta_k(\eta) = \frac{\pi}{2} - \frac{\pi (-k\eta)^{2\nu}}{4^{\nu}\nu \Gamma^2(\nu)},
\end{align}
showing that the leading phase evolution arises from interference between the growing and decaying branches. If only the $C_{\nu}$ branch alone were retained, the phase would freeze to the constant value $\pi/2$, leading to $\theta'_k = 0$ and violating the canonical Wronskian condition $\rho_k^2 \theta_k' =1$.

Using this phase, one finds $\rho_k^2 \theta_k' =1$, which again yields $I_{EL} = 1/2$.
The value $I_{EL} = 1/2$ is set by the Bunch-Davies initial condition and is conserved exactly by the $SL(2,\mathbb{R})$ symmetry throughout the inflationary evolution \cite{Niederer:1973tz}. The slow-roll parameters $\epsilon$ and $\delta$ enter only through $\nu$ and affect the form of the generators $T_1,\,T_2$ and hence the squeezing quantities $I_1,\,I_2$ of Eqs. (\ref{gentdho1}) and (\ref{gentdho2}), but cannot alter $I_{EL} = 2 T_0$.

\section{Summary and Discussions}

In this paper we have investigated the dynamical symmetry structure of the Sasaki-Mukhanov equation governing primordial scalar perturbations during inflation.

The Sasaki-Mukhanov equation maps exactly to a simple harmonic oscillator by introducing the Ermakov auxiliary field $\rho_k(\eta)$ satisfying the nonlinear Ermakov-Pinney equation. The conserved energy of this transformed system is the Ermakov-Lewis invariant $I_{EL}$, an exact nonperturbative constant of motion for any inflationary background. The three generators $(T_0,\,T_1,\,T_2)$ of the Ermakov-transformed system close on the $sl(2,R)$ Lie algebra with $I_{EL} =2 T_0$.

Applied to de Sitter and slow-roll inflation, we find $I_{EL} =1/2$ throughout the entire evolution, in both the sub- and super-horizon regimes. This value is fixed by the canonical normalization of the Bunch-Davies vacuum and is then conserved exactly by the $SL(2, \mathbb{R})$ symmetry. Slow-roll corrections shift the index $\nu=3/2$ to $\nu=3/2 + 2\epsilon -\delta$ and modify the squeezing observables $I_1,\,I_2$, but leave $I_{EL}$ unchanged at any order in slow-roll.

Unlike the approximate super-horizon conservation of the curvature perturbation $\mathcal{R}_k$ in slow-roll inflation, which follows from the suppression of the decaying mode, the Ermakov–Lewis invariant is an exact dynamical invariant of the quadratic Sasaki-Mukhanov system. As a consequence, it provides an exact characterization of inflationary mode evolution even in strongly nonadiabatic regimes where the standard slow-roll and WKB approximations break down. This suggests that the Ermakov–Lewis framework may be particularly useful for studying beyond-slow-roll scenarios such as ultra-slow-roll inflation, transient feature models, and possibly stochastic inflation.



\vspace{5mm}
{\bf Research Funding:}
This work was supported by the research grant of Jeju National University in 2023.

\bibliography{refs}

@article{Niederer:1973tz,
    author = "Niederer, U.",
    title = "{The maximal kinematical invariance group of the harmonic oscillator}",
    reportNumber = "PRINT-72-4208",
    journal = "Helv. Phys. Acta",
    volume = "46",
    pages = "191--200",
    year = "1973"
}

@article{Bunch:1978yq,
    author = "Bunch, T. S. and Davies, P. C. W.",
    title = "{Quantum Field Theory in de Sitter Space: Renormalization by Point Splitting}",
    doi = "10.1098/rspa.1978.0060",
    journal = "Proc. Roy. Soc. Lond. A",
    volume = "360",
    pages = "117--134",
    year = "1978"
}

@article{Stewart:1993bc,
    author = "Stewart, Ewan D. and Lyth, David H.",
    title = "{A More accurate analytic calculation of the spectrum of cosmological perturbations produced during inflation}",
    eprint = "gr-qc/9302019",
    archivePrefix = "arXiv",
    reportNumber = "KUNS-1176, LANCS-TH-93-01",
    doi = "10.1016/0370-2693(93)90379-V",
    journal = "Phys. Lett. B",
    volume = "302",
    pages = "171--175",
    year = "1993"
}

@article{deAlfaro:1976vlx,
    author = "de Alfaro, Vittorio and Fubini, S. and Furlan, G.",
    title = "{Conformal Invariance in Quantum Mechanics}",
    reportNumber = "CERN-TH-2115",
    doi = "10.1007/BF02785666",
    journal = "Nuovo Cim. A",
    volume = "34",
    pages = "569",
    year = "1976"
}

@article{Leach:1977a,
  author  = {Leach, P. G. L.},
  title   = {On the theory of time-dependent linear canonical 
             transformations as applied to {H}amiltonians of the 
             harmonic oscillator type},
  journal = {J. Math. Phys.},
  volume  = {18},
  pages   = {1608--1611},
  year    = {1977},
  doi     = {10.1063/1.523447}
}

@article{Ermakov:1880,
  author  = {Ermakov, V. P.},
  title   = {Second order differential equations: 
             Conditions of complete integrability},
  journal = {Universitetskie Izvestiya Kiev},
  volume  = {9},
  pages   = {1--25},
  year    = {1880},
  note    = {Translated by A. O. Harin, 
             Appl. Anal. Discrete Math. \textbf{2}, 123 (2008)}
}

@article{Lewis:1968tm,
    author = "Lewis, H. R. and Riesenfeld, W. B.",
    title = "{An Exact quantum theory of the time dependent harmonic oscillator and of a charged particle time dependent electromagnetic field}",
    doi = "10.1063/1.1664991",
    journal = "J. Math. Phys.",
    volume = "10",
    pages = "1458--1473",
    year = "1969"
}

@article{Lidsey:1995np,
    author = "Lidsey, James E. and Liddle, Andrew R. and Kolb, Edward W. and Copeland, Edmund J. and Barreiro, Tiago and Abney, Mark",
    title = "{Reconstructing the inflation potential : An overview}",
    eprint = "astro-ph/9508078",
    archivePrefix = "arXiv",
    reportNumber = "SUSSEX-AST-95-8-3, FERMILAB-PUB-95-280-A",
    doi = "10.1103/RevModPhys.69.373",
    journal = "Rev. Mod. Phys.",
    volume = "69",
    pages = "373--410",
    year = "1997"
}

@article{Liddle:1992wi,
    author = "Liddle, Andrew R. and Lyth, David H.",
    title = "{COBE, gravitational waves, inflation and extended inflation}",
    eprint = "astro-ph/9208007",
    archivePrefix = "arXiv",
    reportNumber = "SUSSEX-AST-92-6-1, LANC-TH-5-92",
    doi = "10.1016/0370-2693(92)91393-N",
    journal = "Phys. Lett. B",
    volume = "291",
    pages = "391--398",
    year = "1992"
}

@article{Leach:1977fa,
    author = "Leach, P. G. L.",
    title = "{Invariants and Wave Functions for Some Time Dependent Harmonic Oscillator Type Hamiltonians}",
    doi = "10.1063/1.523161",
    journal = "J. Math. Phys.",
    volume = "18",
    pages = "1902--1907",
    year = "1977"
}

@inproceedings{Pinney:1950,
  title={The nonlinear differential equation y+ p (x) y+ cy- 3= 0},
  author={Pinney, Edmund},
  booktitle={Proc. Amer. Math. Soc},
  volume={1},
  number={681},
  pages={1},
  year={1950}
}

@article{Lewis:1967,
  author = {Lewis, H. R.},
  title = {Classical and quantum systems with time-dependent harmonic-oscillator-type Hamiltonians},
  journal = {Physical Review Letters},
  volume = {18},
  year = {1967},
  pages = {510}
}

@book{Wybourne:1974,
  title={Classical Groups for Physicists},
  author={Wybourne, B.G.},
  isbn={9780471965053},
  lccn={73017363},
  series={A Wiley-Interscience publication},
  url={https://books.google.co.kr/books?id=OQGiieSztuIC},
  year={1974},
  publisher={Wiley}
}

@article{Polarski:1995jg,
    author = "Polarski, David and Starobinsky, Alexei A.",
    title = "{Semiclassicality and decoherence of cosmological perturbations}",
    eprint = "gr-qc/9504030",
    archivePrefix = "arXiv",
    reportNumber = "LMPM-95-4",
    doi = "10.1088/0264-9381/13/3/006",
    journal = "Class. Quant. Grav.",
    volume = "13",
    pages = "377--392",
    year = "1996"
}

@article{Albrecht:1992kf,
    author = "Albrecht, Andreas and Ferreira, Pedro and Joyce, Michael and Prokopec, Tomislav",
    title = "{Inflation and squeezed quantum states}",
    eprint = "astro-ph/9303001",
    archivePrefix = "arXiv",
    reportNumber = "IMPERIAL-TP-92-93-21",
    doi = "10.1103/PhysRevD.50.4807",
    journal = "Phys. Rev. D",
    volume = "50",
    pages = "4807--4820",
    year = "1994"
}

@article{Grishchuk:1990bj,
    author = "Grishchuk, L. P. and Sidorov, Yu. V.",
    title = "{Squeezed quantum states of relic gravitons and primordial density fluctuations}",
    doi = "10.1103/PhysRevD.42.3413",
    journal = "Phys. Rev. D",
    volume = "42",
    pages = "3413--3421",
    year = "1990"
}

@article{Parker:1969au,
    author = "Parker, Leonard",
    title = "{Quantized fields and particle creation in expanding universes. 1.}",
    doi = "10.1103/PhysRev.183.1057",
    journal = "Phys. Rev.",
    volume = "183",
    pages = "1057--1068",
    year = "1969"
}

@article{Mukhanov:1990me,
    author = "Mukhanov, Viatcheslav F. and Feldman, H. A. and Brandenberger, Robert H.",
    title = "{Theory of cosmological perturbations. Part 1. Classical perturbations. Part 2. Quantum theory of perturbations. Part 3. Extensions}",
    reportNumber = "BROWN-HET-796, BROWN-HET-800, BROWN-HET-780",
    doi = "10.1016/0370-1573(92)90044-Z",
    journal = "Phys. Rept.",
    volume = "215",
    pages = "203--333",
    year = "1992"
}

@article{Planck:2018jri,
    author = "Akrami, Y. and others",
    collaboration = "Planck",
    title = "{Planck 2018 results. X. Constraints on inflation}",
    eprint = "1807.06211",
    archivePrefix = "arXiv",
    primaryClass = "astro-ph.CO",
    doi = "10.1051/0004-6361/201833887",
    journal = "Astron. Astrophys.",
    volume = "641",
    pages = "A10",
    year = "2020"
}

@article{Linde:1981mu,
    author = "Linde, Andrei D.",
    editor = "Fang, Li-Zhi and Ruffini, R.",
    title = "{A New Inflationary Universe Scenario: A Possible Solution of the Horizon, Flatness, Homogeneity, Isotropy and Primordial Monopole Problems}",
    reportNumber = "LEBEDEV-81-229",
    doi = "10.1016/0370-2693(82)91219-9",
    journal = "Phys. Lett. B",
    volume = "108",
    pages = "389--393",
    year = "1982"
}

@article{Guth:1980zm,
    author = "Guth, Alan H.",
    editor = "Fang, Li-Zhi and Ruffini, R.",
    title = "{The Inflationary Universe: A Possible Solution to the Horizon and Flatness Problems}",
    reportNumber = "SLAC-PUB-2576",
    doi = "10.1103/PhysRevD.23.347",
    journal = "Phys. Rev. D",
    volume = "23",
    pages = "347--356",
    year = "1981"
}

@article{Starobinsky:1979ty,
    author = "Starobinsky, Alexei A.",
    editor = "Khalatnikov, I. M. and Mineev, V. P.",
    title = "{Spectrum of relict gravitational radiation and the early state of the universe}",
    journal = "JETP Lett.",
    volume = "30",
    pages = "682--685",
    year = "1979"
}

@article{Mukhanov:1988jd,
    author = "Mukhanov, Viatcheslav F.",
    title = "{Quantum Theory of Gauge Invariant Cosmological Perturbations}",
    journal = "Sov. Phys. JETP",
    volume = "67",
    pages = "1297--1302",
    year = "1988"
}

@article{Sasaki:1986hm,
    author = "Sasaki, Misao",
    title = "{Large Scale Quantum Fluctuations in the Inflationary Universe}",
    reportNumber = "RRK-86-29",
    doi = "10.1143/PTP.76.1036",
    journal = "Prog. Theor. Phys.",
    volume = "76",
    pages = "1036",
    year = "1986"
}

@article{Kodama:1984ziu,
    author = "Kodama, Hideo and Sasaki, Misao",
    title = "{Cosmological Perturbation Theory}",
    doi = "10.1143/PTPS.78.1",
    journal = "Prog. Theor. Phys. Suppl.",
    volume = "78",
    pages = "1--166",
    year = "1984"
}

@article{Bardeen:1980kt,
    author = "Bardeen, James M.",
    title = "{Gauge Invariant Cosmological Perturbations}",
    doi = "10.1103/PhysRevD.22.1882",
    journal = "Phys. Rev. D",
    volume = "22",
    pages = "1882--1905",
    year = "1980"
}
\bibliographystyle{unsrt}

\end{document}